\documentclass{JCIN22}

\usepackage[sort&compress]{gbt7714}

\usepackage{amsmath,amsfonts}
\usepackage{array}
\usepackage{textcomp}
\usepackage{stfloats}
\usepackage{url}
\usepackage{verbatim}
\usepackage{amssymb}
\usepackage{array}
\usepackage{amsfonts}
\usepackage{amsmath}
\usepackage{amsthm}
\usepackage{mathrsfs}
\usepackage{bm}
\DeclareMathOperator{\diag}{diag}

\renewcommand{\mathbf}[1]{\bm{#1}}
\renewcommand{\pi}{\uppi}

\newcommand{\herm}{\mathrm{H}}
\newsavebox{\slantedgreekbox}
\newcommand{\slantedbmaux}[2]{%
  \mbox{%
    \sbox{\slantedgreekbox}{$#1\bm{#2}$}%
    \hskip\wd\slantedgreekbox
    \pdfsave
    \pdfsetmatrix{1 0 0.25 1}%
    \llap{\usebox{\slantedgreekbox}}%
    \pdfrestore
  }%
}
\newcommand{\slantedbm}[1]{%
  \mathchoice
    {\slantedbmaux{\displaystyle}{#1}}%
    {\slantedbmaux{\textstyle}{#1}}%
    {\slantedbmaux{\scriptstyle}{#1}}%
    {\slantedbmaux{\scriptscriptstyle}{#1}}%
}
\usepackage{verbatim}
\usepackage{autobreak} 
\usepackage{lettrine}
\usepackage{float} 
\usepackage{tabularx}
\usepackage{enumerate}
\usepackage{cases}
\usepackage{url}
\usepackage[colorlinks,
            linkcolor=black,
            anchorcolor=black,
            citecolor=blue]{hyperref}
\usepackage{orcidlink}
\usepackage{hyperref}
\usepackage{graphicx}
\usepackage{caption} 
\newcommand{\inlinecite}[1]{{\setcitestyle{numbers,square,comma}\cite{#1}}}

\begin{document}

\ArticleType{}
\Year{2022}

\title{A New Paradigm Towards Reconfigurable Environment: Reconfigurable Distributed Antennas and Reflecting Surface}

\author[]{Jintao Wang}
\author[]{Pingping Zhang}
\author[]{Chengzhi Ma}
\author[]{Chengwang Ji}
\author[]{Zheng Shi}
\author[]{Guanghua Yang}
\author[]{Shaodan Ma}

\maketitle

\barefootnote{

Jintao Wang and Chengzhi Ma are with the State Key Laboratory of Internet of Things for Smart City, University of Macau, Macau 999078, China, and also with the School of Intelligent Systems Science and Engineering, Jinan University, Zhuhai 519070, China. (email: wang.jintao@connect.um.edu.mo; yc07499@um.edu.mo). \\\indent
Pingping Zhang is with the School of Electronics and Information Engineering, South China Normal University, Foshan 528000, China (e-mail: pingpingzhang@m.scnu.edu.cn).\\\indent
Chengwang Ji and Shaodan Ma are with the Department of Electrical and Computer Engineering, and State Key Laboratory of Internet of Things for Smart City, University of Macau, Macau 999078, China (e-mail: ji.chengwang@connect.um.edu.mo; shaodanma@um.edu.mo). \\\indent
Zheng Shi and Guanghua Yang are with the School of Intelligent Systems Science and Engineering, Jinan University, Zhuhai 519070, China (e-mail: zhengshi@jnu.edu.cn; ghyang@jnu.edu.cn).
}

{\bf\textit{Abstract---}Reconfigurable distributed antennas and reflecting surface (RDARS) has emerged as a transformative solution to address the stringent requirements of future wireless networks. By combining distributed active antennas with reconfigurable passive reflecting surfaces, RDARS integrates the advantages of both active transmission and passive wave control in a cost-effective and energy-efficient manner. This hybrid architecture enables enhanced coverage, improved spectral efficiency, and seamless support for integrated communication and sensing. In this article, we first introduce the fundamental architecture and working principles of RDARS, followed by practical benefits and comparisons with recently proposed intelligent surface variants. We then highlight the signal-to-noise ratio (SNR) gains in representative applications of RDARS-aided communication and sensing scenarios, where RDARS demonstrates clear advantages over conventional reconfigurable intelligent surfaces. Finally, we outline key challenges related to practical implementation and resource allocation, and discuss potential research directions. With its unique hybrid mode synergy, RDARS is envisioned to play a pivotal role in shaping the evolution of next-generation intelligent communication systems.\\[-1.5mm]

}


\section{\MakeUppercase{Introduction}}
The sixth generation (6G) of wireless communications is expected to support unprecedented requirements in terms of connectivity, reliability, spectrum efficiency (SE), and energy efficiency (EE). Among the enabling technologies, extremely large-scale multiple-input multiple-output (XL-MIMO) has emerged as a key solution by exploiting spatial diversity and multiplexing gains through a large number of antennas at the base station \cite{10379539}. In combination with radio-over-fiber (RoF) technologies \cite{9409616}, massive MIMO antennas can be deployed either in a centralized or distributed manner, with the latter forming a distributed antenna system (DAS) \cite{7349255,9206553,Moerman2022DAS}. Distributed deployment further leverages rich scattering environments to provide additional distribution gains, making it a promising paradigm for future wireless systems. However, to sustain high SE under stringent quality-of-service (QoS) demands, massive MIMO and DAS architectures typically require a large number of active antennas and radio-frequency (RF) chains, resulting in substantial hardware cost and power consumption.

To alleviate these issues, reconfigurable intelligent surfaces (RISs), also known as intelligent reflecting surfaces (IRSs), have attracted significant attention \cite{9326394,9424177,9206044,9847080}. A RIS is typically composed of a large array of nearly passive elements capable of imposing independent phase shifts on incident signals, thus enabling the reconfiguration of the wireless environment at low cost and with negligible power consumption \cite{10480438}. By intelligently tuning these elements, RIS-assisted networks gain an additional degree of freedom (DoF) to improve link reliability and enhance system efficiency. Nevertheless, RIS deployment in practice faces a fundamental limitation, known as the “multiplicative fading effect”. Performance gains are significant only when the direct transmission link is blocked or severely attenuated, which limits the practical benefits in diverse scenarios \cite{2active2023}.


Recently, several variants of RIS have been proposed to address the aforementioned challenges. For instance, the authors in Ref.\inlinecite{2active2023} and \inlinecite{3active2021} introduced the concept of active RIS, in which each reflecting element is equipped with a power amplifier (PA) to simultaneously amplify the power of the incident signal and adjust its phase, thereby achieving additional amplification gain. However, equipping every unit with a PA incurs high hardware cost and energy consumption. Moreover, the performance of active RIS is not always superior to that of passive RIS due to noise amplification \cite{active_vs_pass_2,active_vs_pass_1}. Specifically, Ref.\inlinecite{active_vs_pass_1} demonstrated that, under limited power budgets and a large number of reflecting elements, a passive RIS can achieve a higher system rate than an active RIS, both of which are deployed with an optimal placement strategy. These insightful results reveal that the advantages of passive and active RIS are complementary rather than mutually exclusive.  

To leverage these complementary benefits, a hybrid active–passive RIS architecture has been proposed, which integrates two co-located sub-surfaces comprising a certain number of passive and active elements \cite{10235893,Peng2023active}. This hybrid design effectively balances the trade-off between performance enhancement and energy efficiency. 
On the other hand, researchers have developed the hybrid-relay RIS (HR-RIS), in which only few elements are connected to RF chains, thereby functioning as relay components capable of signal amplification to significantly enhance system performance \cite{nguyenHybridRelayReflectingIntelligent2022,9685434,10507738,Nguyen2023Spectral}. In Ref.\inlinecite{nguyenHybridRelayReflectingIntelligent2022}, two HR-RIS architectures were proposed, namely the fixed HR-RIS, where the active elements are predetermined during manufacturing, and the dynamic HR-RIS, in which the active elements can be adaptively configured to improve performance while reducing power consumption. 
Nevertheless, the incorporation of additional PAs or RF chains in these advanced RIS architectures not only undermines the ultra-low-power design objectives envisioned for 6G networks but also introduces hardware complexity, thereby posing new challenges for their practical implementation. 
{\color{black}Furthermore, the control mechanisms for RIS variants present distinct challenges, necessitating intelligent controllers for element configuration.}

\begin{figure*}[t]
    \centering
    \subfloat[Ubiquitous Coverage] {\includegraphics[width=.30\textwidth]{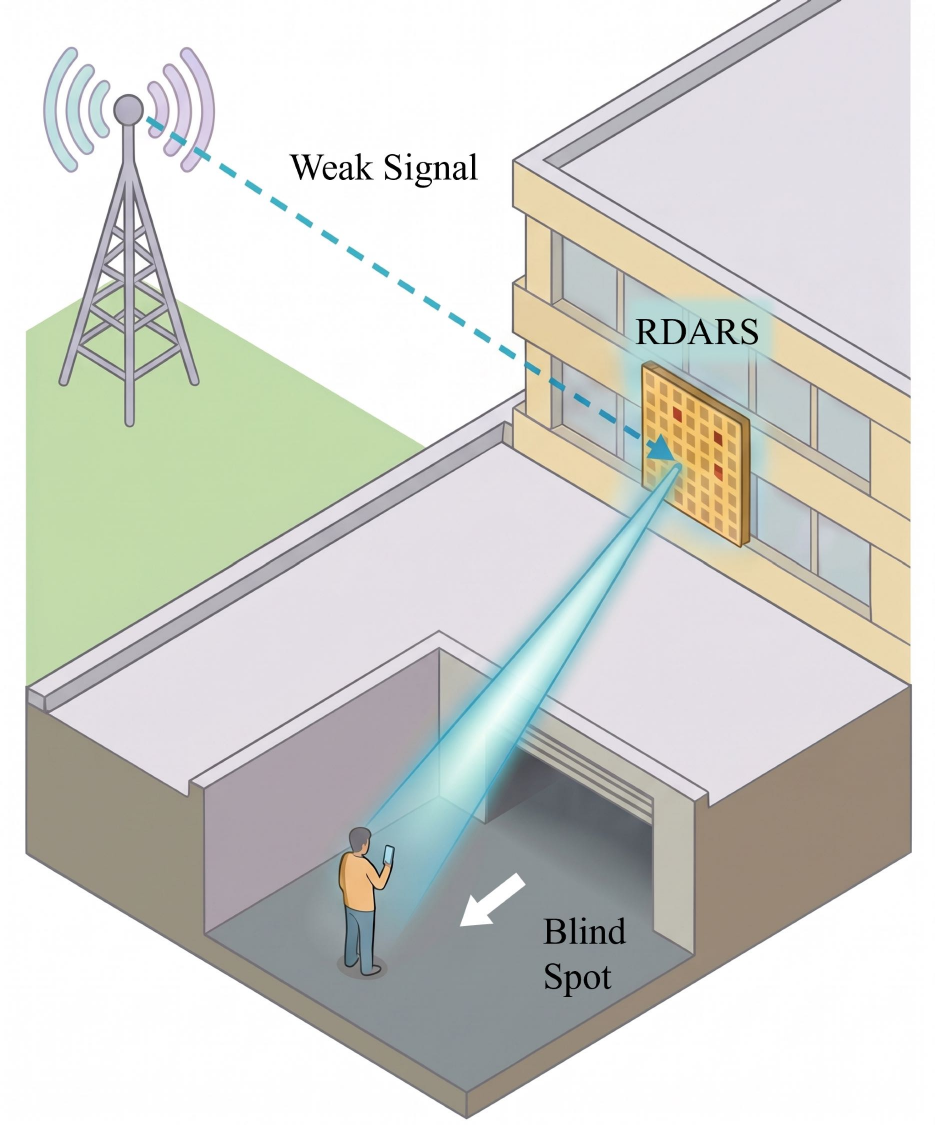}}
    \subfloat[Immersive Smart Transportation] {\includegraphics[width=.30\textwidth]{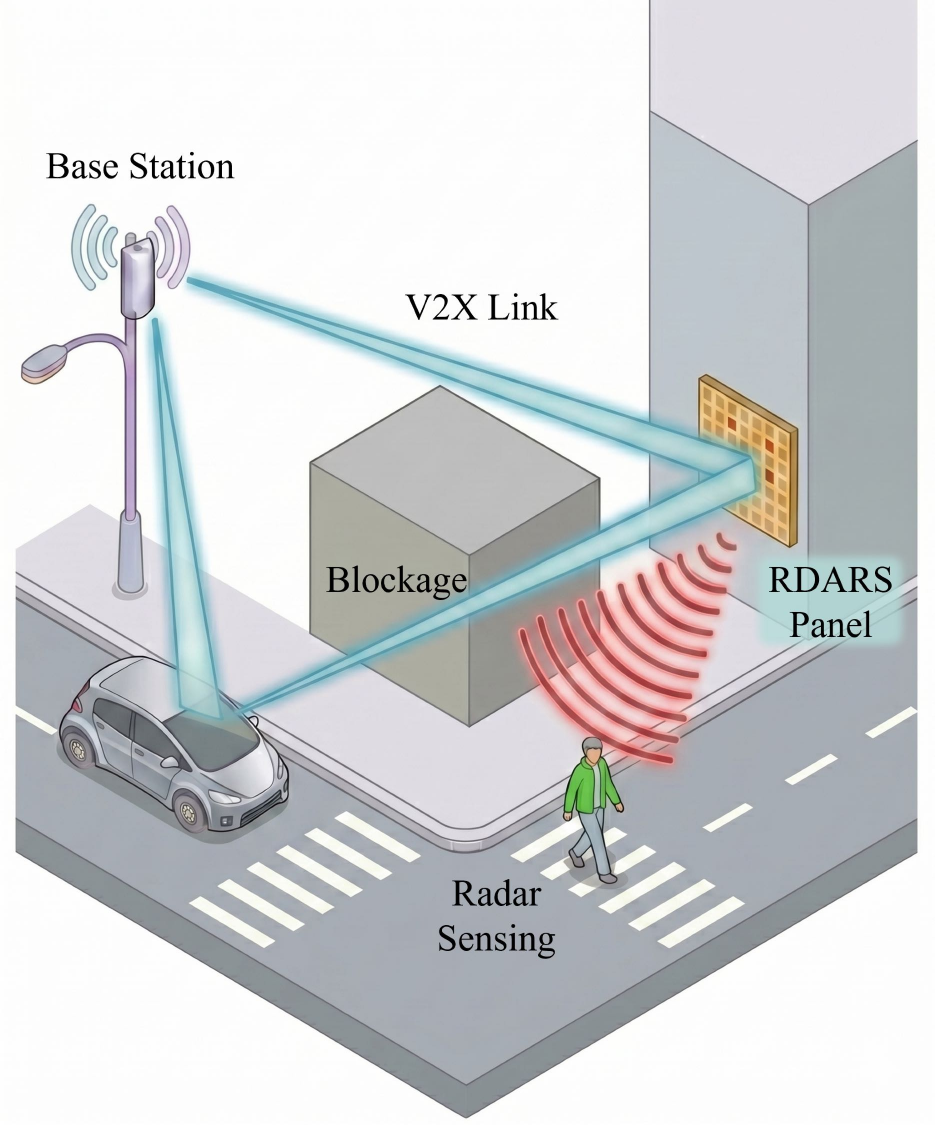}}
    \subfloat[Emergency UAV Networks] {\includegraphics[width=.30\textwidth]{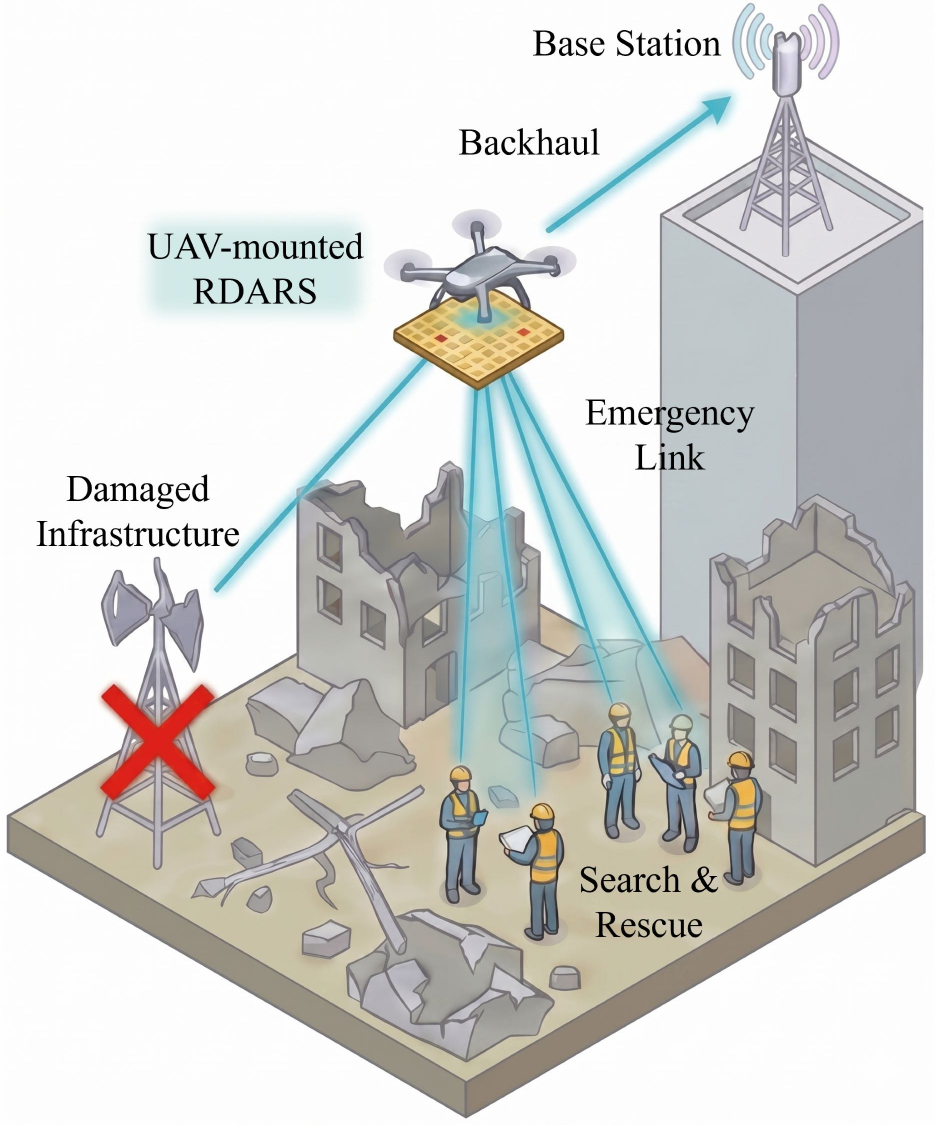}}
    \caption{Typical applications of RDARS-aided wireless communications, including (a) Ubiquitous Coverage (b) Immersive Smart Transportation and (c) Emergency UAV Networks.}
    \label{Fig.app}
\end{figure*}

To overcome these limitations, the reconfigurable distributed antenna and reflecting surface (RDARS) architecture has been recently proposed as a novel paradigm towards reconfigurable
environment for 6G wireless networks \cite{ma2023reconfigurable,wang2024radars,wang2024radarsproceeding,ma2025radars,pei2025radars,zhang2025radars}. RDARS effectively integrates the advantages of both DAS and RIS. It retains the distribution gain of active antennas while leveraging the low-cost passive reflection capability of RIS. Uniquely, each RDARS element can be configured into two operational modes: (i) {\textbf{reflection mode}}, where elements act as passive reflectors similar to RIS, and (ii) {\textbf{connection mode}}, where elements function as distributed antennas to actively transmit or receive signals.\footnote{ \color{black} Note that the RDARS concept is not limited to these two operating modes. For example, it can also be programmed into an absorption mode by terminating the RF ports with negative loads to suppress signal reflections, thereby achieving electromagnetic stealth or facilitating energy harvesting \cite{Patent_RDARS}. In this sense, RDARS can be regarded as a generalization of the multi-state RIS concept \cite{10507738,11184849}, offering a versatile platform that supports both active and passive operations.}
 This dual functionality allows RDARS to inherit the low-cost advantage of RIS while mitigating the multiplicative fading effect. Furthermore, the utilization of connection elements facilitates the control of the RDARS architecture.
In contrast to conventional massive MIMO systems that rely on large-scale deployment of active antennas and RF chains, RDARS strategically enables flexible switching between reflection and connection modes, thereby simultaneously exploiting reflection, distribution, and selection gains.  Crucially, while dynamic HR-RIS is restricted to signal amplification, RDARS leverages connection elements to perform active signal processing. Consequently, the architecture supports cost-efficient upgrades by reusing existing distributed antenna infrastructure, making it highly attractive for practical deployment in 6G networks.

The unique integration of reflection and connection functionalities in RDARS enables its deployment across a wide range of practical scenarios. {\color{black}Several representative applications are illustrated in Fig.~\ref{Fig.app} and discussed as follows.

\begin{itemize} 
\item \textit{Ubiquitous Coverage:}
Unlike conventional RIS, which suffers from double-fading attenuation, RDARS offers a superior solution for deep coverage in scenarios such as cell edges, indoor hotspots, or underground tunnels. In these distinct regions, the connection mode allows RDARS to operate as a low-cost distributed antenna, actively receiving weak signals. This hybrid capability creates robust ``signal hotspots" and effectively eliminates blind spots where purely passive reflection would fail due to severe path loss.

\item \textit{Immersive Smart Transportation:} RDARS is pivotal for enabling high-reliability Vehicle-to-Everything (V2X) networks. In reflection mode, RDARS establishes virtual Line-of-Sight (LoS) links to bypass blockages caused by buildings or large vehicles, ensuring continuous connectivity. More importantly, in connection mode, RDARS elements function as distributed radar sensors to actively capture vehicle echoes. This enables the system to perform collaborative sensing at the roadside.

\item \textit{Emergency UAV Networks:}
RDARS provides a flexible air–ground interface for post-disaster relief and low-altitude economy. In emergency scenarios where terrestrial infrastructure is compromised, RDARS nodes mounted on UAVs can be rapidly deployed. The connection mode ensures reliable control links and facilitates active sensing to locate survivors or map debris. Simultaneously, the reflection mode extends the service range of remaining active base stations, forming a resilient, self-healing mesh network that supports both search-and-rescue operations and temporary data backhaul.

\end{itemize} }

{\color{black}
In this paper, we first present the RDARS architecture and reveal its unique performance gains enabled by dynamic mode configuration. Building on these insights, we then investigate representative RDARS-aided communication and ISAC systems, followed by a discussion of the main practical challenges and promising future research directions. 

Different from \inlinecite{wang2024radars}, which mainly focuses on the joint beamforming optimization and mode selection problem for RDARS-assisted MIMO systems, this paper aims to provide a broader and more systematic overview of the RDARS paradigm from the perspectives of architecture, performance gains, representative applications, and challenges. Specifically, the main contributions of this paper are summarized as follows:
\begin{itemize}
    \item We provide a unified introduction to the RDARS architecture and its hardware operating principles.
    \item We present a structured discussion of the practical advantages of RDARS, together with comparisons to related intelligent surface architectures.
    \item We offer a cross-scenario interpretation of the key performance gains enabled by RDARS in both communication and ISAC systems.
    \item We discuss practical implementation issues, resource allocation challenges, and directions toward future cell-free RDARS networks.
\end{itemize}

The remainder of this paper is organized as follows. Section 2 introduces the RDARS architecture, including its working principle, performance gains, practical benefits, and comparisons with related structures. Section 3 and Section 4 investigate RDARS-aided communication and ISAC systems, respectively. Section 5 discusses open challenges and future research directions. Finally, Section 6 concludes this paper.
}


\begin{figure}[t]
    \centering
    \includegraphics[width=0.45\textwidth]{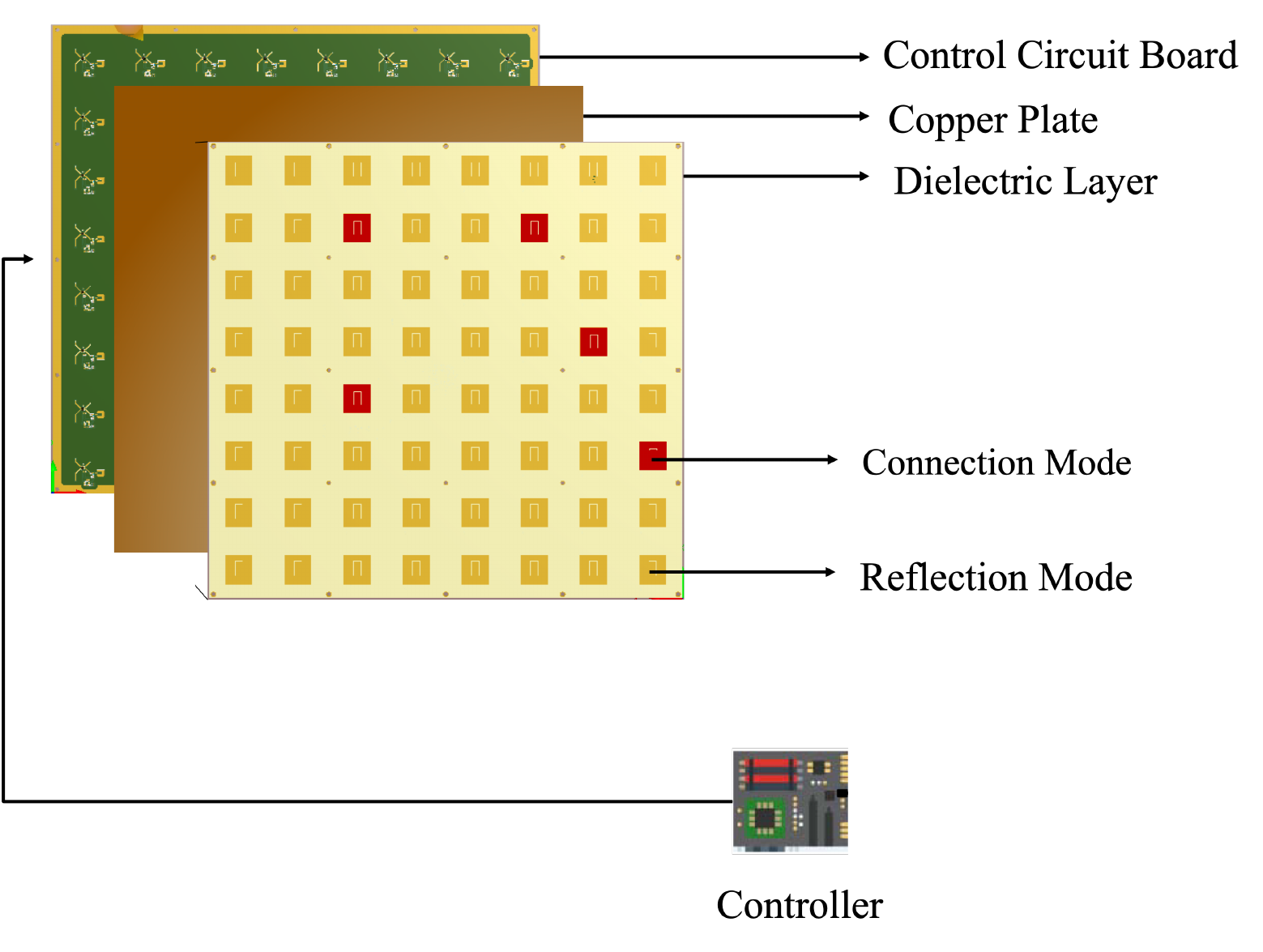}
    \caption{Hardware architecture of RDARS.}
    \label{Fig.architure} 
\end{figure}

\section{\MakeUppercase{RDARS Architecture}}
In this section, we first introduce the RDARS architecture in terms of its working principle, performance gains, practical benefits and comparison with other structures, which together highlight the key advantages of this emerging paradigm.

\subsection{Hardware Architecture} 
As illustrated in Fig.~\ref{Fig.architure}, the RDARS architecture comprises a central controller and a planar array of reconfigurable elements, each capable of operating in two distinct modes: connection and reflection. In the reflection mode, the elements function as passive reflectors that impose programmable phase shifts on incident signals, thereby enabling reconfiguration of the wireless propagation environment in a manner analogous to a conventional RIS. In contrast, in the connection mode, the elements serve as distributed antennas directly connected to the base station (BS) through dedicated optical fibers or wired fronthaul links, thus providing the distribution gain characteristic of a DAS. These connection elements are capable of performing signal transmission and reception for both downlink and uplink communications. Owing to their inherent working principle, no RF chains are required for signal processing at the RDARS, thereby preserving the low-hardware complexity advantage of conventional RISs. The coexistence of these two modes enables RDARS to flexibly strike a balance between low-cost reconfigurability and reliable active transmission.

\begin{figure}[t]
    \centering
    \includegraphics[width=0.4\textwidth]{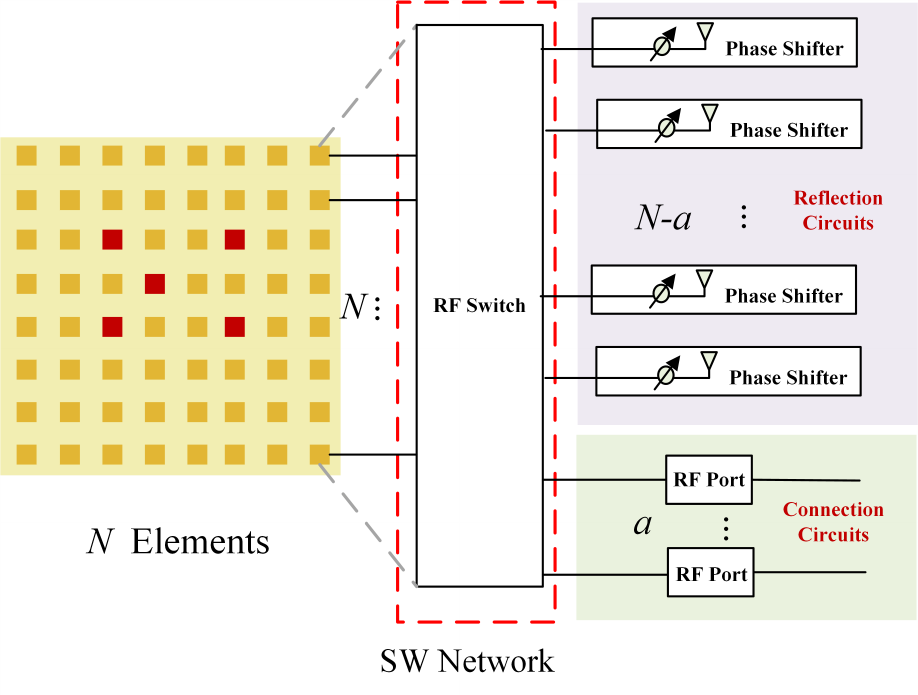}
    \caption{Illustration of RDARS working principle {\color{black}\protect\inlinecite{wang2024radars}}.}
    \label{Fig.Con} 
\end{figure}

\begin{figure}[t]
    \centering
    \subfloat[Front view] {\includegraphics[width=.20\textwidth]{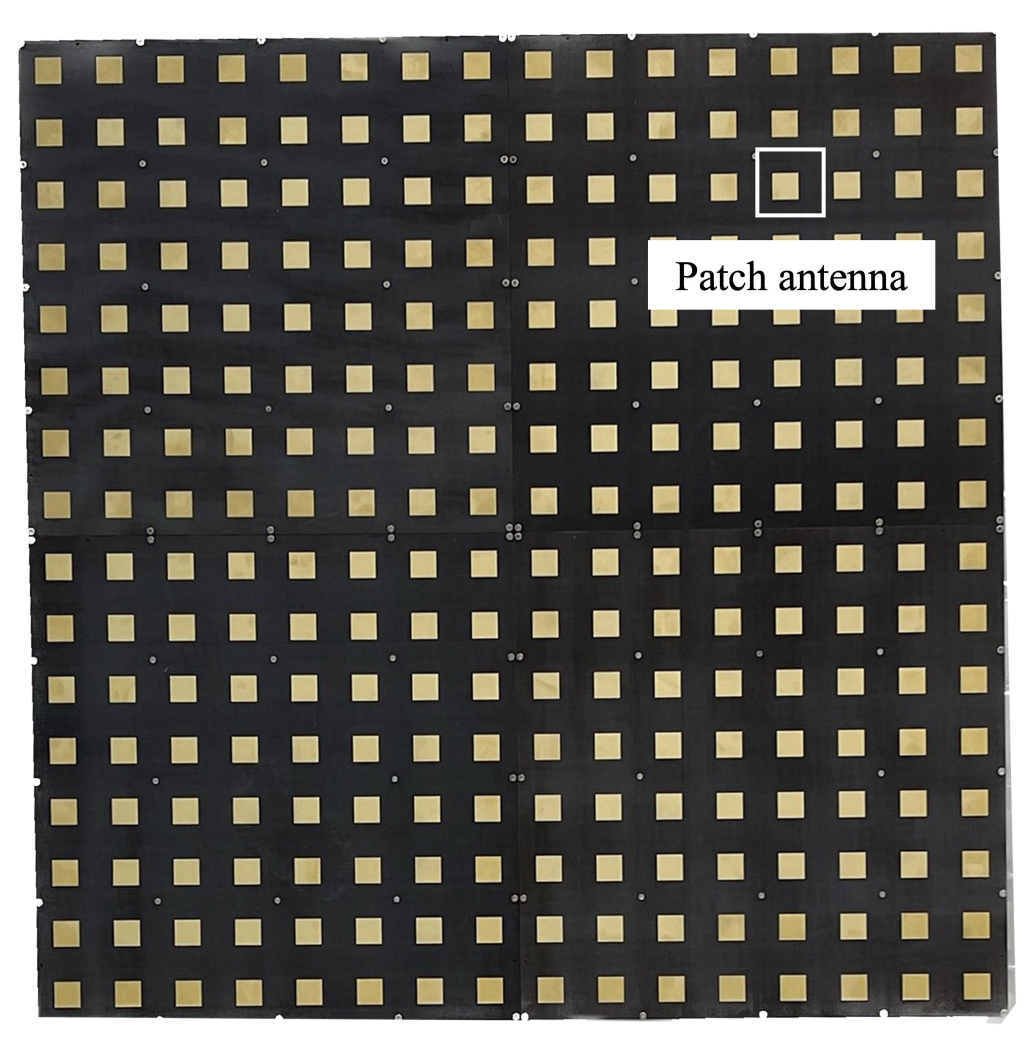}}
    \hspace{-6pt}
    \subfloat[Rear view] {\includegraphics[width=.20\textwidth]{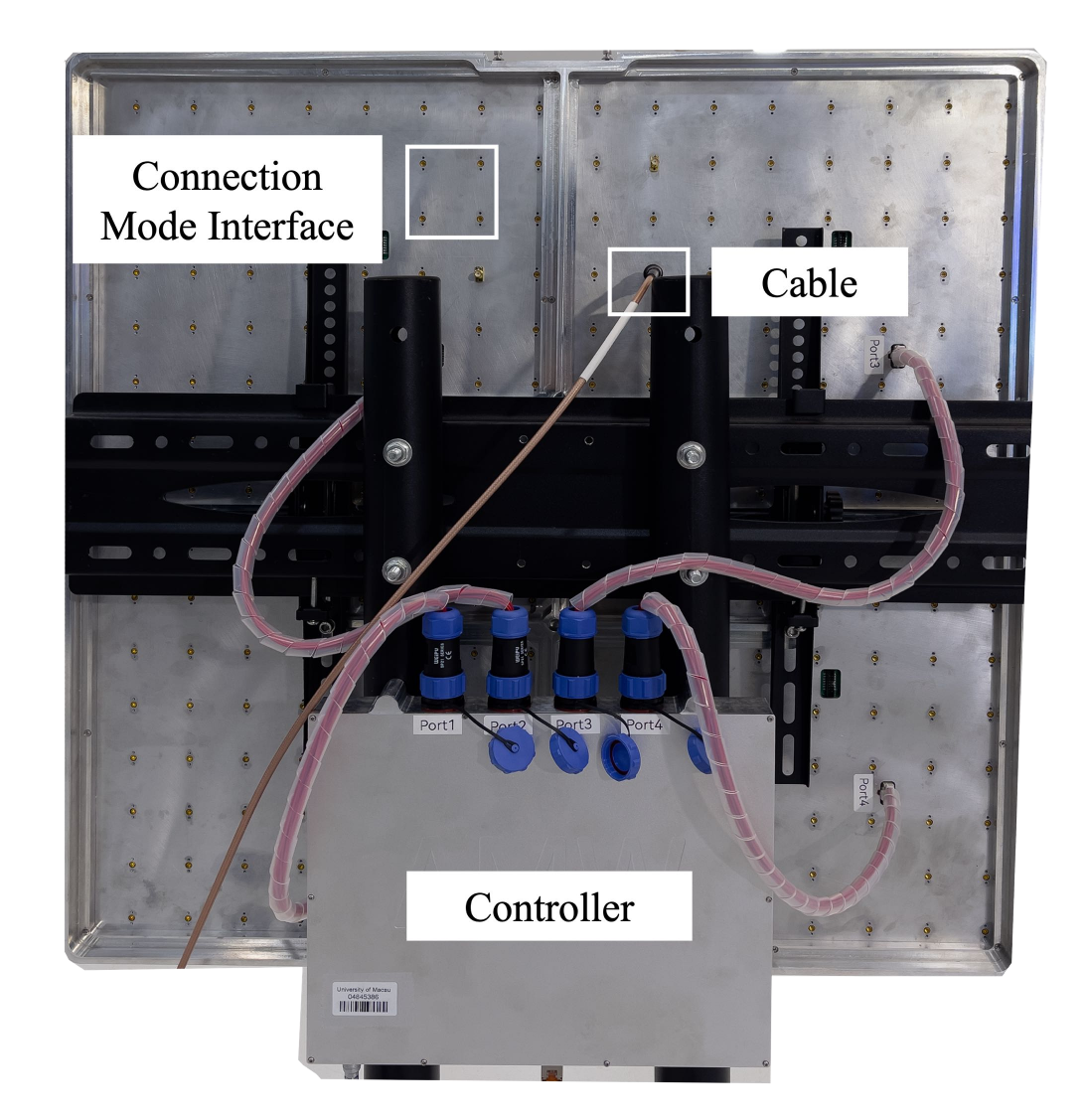}}
    \caption{Illustration of the RDARS prototype with $16\times16$ elements.}
    \label{fig_prototype}
\end{figure}

Illustration of RADRS operating principle is provided in Fig~\ref{Fig.Con}. Each element can be programmed into either connection or reflection mode, providing switching DoFs that enable diversity enhancement. The two operating modes are realized through their respective connection and reflection circuits. 
The dynamic switching between these two modes is achieved via a switching network (SW), typically implemented using RF or complementary metal-oxide-semiconductor (CMOS) switches \cite{10507738,Gao2018Hybrid}. A fabricated prototype of RDARS with $16\times16$ elements is illustrated in Fig.~\ref{fig_prototype}. 

Consequently, a RDARS encompasses passive RIS and DAS as two extreme cases corresponding to the full-reflection and full-connection configurations, respectively.
Recent studies have demonstrated that even a small subset of elements configured in the connection mode can yield substantial performance improvements without requiring a large number of active circuits~\cite{ma2023reconfigurable}. Motivated by this observation, the RDARS architecture employs only $a$ connection circuits, while the remaining $N - a$ elements operate in the reflection mode, where $N$ denotes the total number of elements and $a \ll N$. This design achieves an effective trade-off among system complexity, hardware cost, and achievable performance, rendering it a highly practical and scalable solution for large-scale deployments.  

    \subsection{Performance Gains}
    The distinctive architecture of RDARS enables multiple forms of performance gains that cannot be simultaneously achieved by other intelligent surfaces.  Specifically, three major gains characterize its performance advantages in 6G wireless networks, detailed as
    \begin{itemize}
    \item {\textit{Reflection gain}}: Reflection gain arises from elements operating in the passive reflection mode, wherein programmable phase shifts are applied to incident signals to provide configurable reflective links. Such behavior resembles that of conventional passive RISs while maintaining low power consumption.

    \item {\textit{Distribution gain}}: Distribution gain is contributed by elements operating in the connection mode, which act as distributed transmit or receive antennas. By establishing direct links to the BS, these elements effectively mitigate the multiplicative fading limitation inherent to purely passive RISs.

    \item {\textit{Selection gain}}: Selection gain is achieved through the dynamic switching between the connection and reflection modes. The programmable switching mechanism introduces additional DoFs, allowing RDARS to optimize mode allocation in response to varying propagation conditions and to realize multiple system functionalities.
    \end{itemize}

    \subsection{Practical Benefits}
    For the practical deployment, RDARS benefits from its ability to reuse existing infrastructure.  Specifically, the fronthual of the distributed antennas in RDARS  can be realized through mature optical fronthaul solutions, such as radio-frequency-overfiber (RFoF) \cite{Novak2016RFoF}, ensuring the integration with existing optical infrastructures.  It can be seamlessly integrated with architectures such as cloud radio access networks (C-RAN), ultra-dense networks (UDNs), and cell-free massive MIMO (CF-mMIMO), thereby reducing deployment complexity and cost. On the other hand, RDARS requires intelligent controllers to configure the elements as in RIS systems. Such control signaling can be realized through wired connections, IP routing, or wireless interfaces \cite{RISwhitepaper2022}. While wireless control provides higher flexibility, it also introduces challenges such as precise synchronization, interference with access links, and limited beamforming accuracy. Fortunately, the utilization of connection elements facilitates the control of RDARS  via the integrated frauthauls.

\subsection{Comparisons}
To further contextualize RDARS within the landscape of intelligent surfaces, we compare it with the recent proposed architectures of intelligent surfaces in terms of structural design, operational flexibility, and performance capabilities.

\subsubsection{RDARS vs Hybrid Active-Passive RIS \cite{10235893}}
The concept of active–passive RIS originates from the active RIS, whose elements are fully equipped with active loads or negative resistances, such as tunnel diodes or negative impedance converters. Each active element is powered by an external supply to enable signal amplification. Consequently, an active RIS can not only provide adjustable phase shifts, as in the conventional passive RIS, but also amplify the amplitude of the incident signals in a full-duplex manner. Building upon this concept, the active–passive RIS architecture incorporates a limited number of active elements into a conventional passive array, enabling simultaneous signal reflection and amplification while maintaining low energy consumption. {\color{black}However, this design lacks the flexibility to dynamically switch between active and passive modes and suffers from the inherent drawback of noise amplification.}
In contrast, RDARS combats multiplicative fading by connection elements and allows each element to switch between reflection and connection modes. As a result, RDARS intelligently activates only the most beneficial elements, achieving higher distribution gain and avoiding unnecessary energy consumption.

\subsubsection{RDARS vs HR-RIS \cite{nguyenHybridRelayReflectingIntelligent2022}}
HR-RIS, also known as semi-passive RIS \cite{9685434}, integrates passive reflecting and active relaying functionalities by activating a small subset of RIS elements with RF chains and amplifiers. This architecture originates from the performance gap observed between conventional RIS and relay systems, where pure passive reflection limits beamforming flexibility. By combining the low-power reflection of RIS and the signal amplification of relays, HR-RIS effectively enhances system performance with only a few active elements, compensating for hardware constraints such as limited phase resolution.
Nevertheless, although HR-RIS enhances the performance of conventional RIS by introducing active relaying elements, it suffers from higher energy consumption, noise amplification, and increased implementation complexity due to the incorporation of additional RF chains. In contrast, RDARS integrates distributed active antennas with passive reflecting surfaces, achieving superior energy efficiency, scalability, and robustness, and thus represents a more promising architecture for future large-scale and low-power wireless systems.

\begin{figure}[t]
    \centering
    \includegraphics[width=0.45\textwidth]{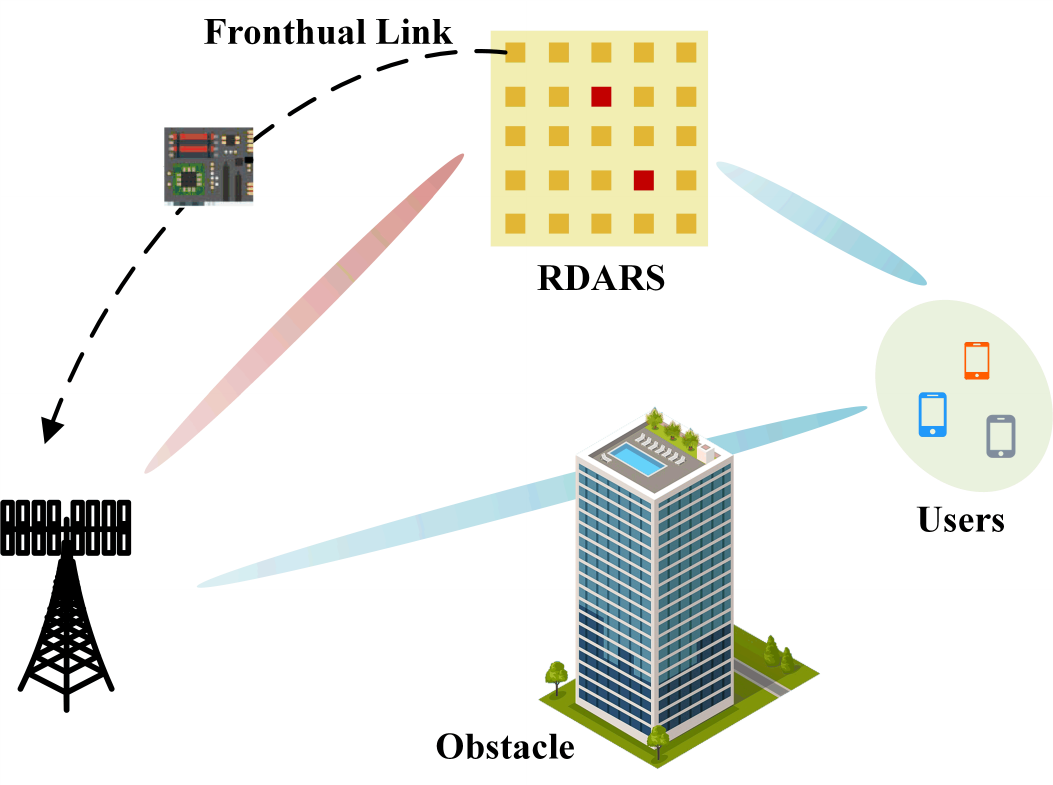}
    \caption{Illustration of RDARS-aided uplink transmission {\color{black}\protect\inlinecite{wang2024radars}}.}
    \label{Fig.updp} 
\end{figure}

\section{\MakeUppercase{RDARS-aided Communications}}\label{sec:rdars_aided_communications}
In this section, we present the system model of the RDARS-assisted communication framework, taking uplink transmission as an example. As shown in Fig.~\ref{Fig.updp}, a RDARS-aided multi-user communication system with single-antenna users and one BS equipped with $M$ antennas is considered. Moreover, we consider the time division multiple access (TDMA) scheme, where the users are served by the BS over orthogonal time slots.  The RDARS is composed of $N$ elements, with $a$ elements operating in the connection mode, while the remaining $N\!-\!a$ passive elements operate in the reflection mode. 

Denoted by $\mathcal{A}$ the index set of the connected components, the mode selection matrix can be represented by the diagonal matrix $\mathbf{A}\in \mathbb{R}^{N \times N}$, whose diagonal value equal to 1 for the connection mode and 0 for the reflection mode. 
Suppose ${\bm{\theta}}\in \mathbb{C}^{N \times 1}$ denotes the virtual reflection coefficient vector of all elements in the RDARS, and define ${\slantedbm{\Phi}}\triangleq\diag({\bm{\theta}})$. Then, the reflection coefficients corresponding to the elements operating in the reflection mode can be expressed as $(\mathbf{I} - \mathbf{A}){\bm{\theta}} \in \mathbb{C}^{N \times 1}$.
%

\subsection{System Model}
In the RDARS-aided uplink communication system, $a$ connection elements in RDARS and BS antennas are expected to receive signals from the user.
Considering the quasi-static far-field channel environment,  the channel from the user to the BS and RDARS can be denoted as $\mathbf{h}_{d} \in\mathbb{C}^{M \times 1}$ and $\mathbf{h}_{r} \in \mathbb{C}^{N \times 1}$, respectively.
 In addition, the channel between the BS and the RDARS is represented as $\mathbf{G} \in \mathbb{C}^{N \times M}$.
Denoted by ${{s}}$ the transmitted symbols from the user with transmit power $\mathbb{E}[{{s}} {{s}}^*]=p$, the uplink received signal $\mathbf{y}\in \mathbb{C}^{(M+a) \times 1}$ at the BS can be divided into two parts, i.e., $\mathbf{y}_{u}^{b}\in \mathbb{C}^{M \times 1}$ and $\mathbf{y}_{u}^{c}\in \mathbb{C}^{a \times 1}$, defined as follows:
\begin{equation}
  \underbrace{\begin{bmatrix} \mathbf{y}_{u}^{b} \\ \mathbf{y}_{u}^{c} \end{bmatrix}}_{\mathbf{y}_{u}}
  = 
  \underbrace{\begin{bmatrix} \mathbf{h}_{d} \!+\!\mathbf{G}^{\herm} (\mathbf{I}\!-\!\mathbf{A}) {\slantedbm{\Phi}} \mathbf{h}_{r} \\  \mathbf{A}_{a} \mathbf{h}_{r}  \end{bmatrix}   }_{ \mathbf{h} } {{s}} + \underbrace{\begin{bmatrix} \mathbf{n}_{b} \\  \mathbf{n}_{c} \end{bmatrix} }_{\mathbf{n}_{u}},
\end{equation} 
where $\mathbf{y}_{u}^{b}$ and $\mathbf{y}_{u}^{c}$ represent the signals received by the BS's antennas and the RDARS's elements operating in the connection mode, respectively.  $\mathbf{n}_{b}$ and $\mathbf{n}_{c}$ represent the additive white Gaussian noise (AWGN) obeying the Circularly Symmetric Complex Gaussian (CSCG) distributions with zero mean and covariance matrices ${\slantedbm{\Lambda}}_{b} \!=\! {\sigma_{b}^2}\mathbf{I}_{M}$ and ${\slantedbm{\Lambda}}_{c} \!=\!{\sigma_{c}^2}\mathbf{I}_{a}$, i.e., $\mathbf{n}_{b} \sim \mathcal{CN}(\mathbf{0},{\sigma_{b}^2}\mathbf{I}_{M})$ and $\mathbf{n}_{c} \sim \mathcal{CN}(\mathbf{0},{\sigma_{c}^2}\mathbf{I}_{a})$, respectively. For simplicity, we assume ${\sigma_{b}^2}={\sigma_{c}^2}={\sigma^2}$ \cite{1545835}.

%

The effective channel $\mathbf{h}$ is composed of two parts, i.e., $\mathbf{h}_{b}$ and $\mathbf{h}_{c}$. 
First, $\mathbf{h}_{b}$ represents the effective channel between the user and the BS assisted with the RDARS's elements in the reflection mode, similar to that of the conventional RIS-aided system, defined as $\mathbf{h}_{b}=\mathbf{h}_{d}+\mathbf{G}^{\herm} (\mathbf{I}\!-\!\mathbf{A}) {\slantedbm{\Phi}} \mathbf{h}_{r}$.
On the other hand, $\mathbf{h}_{c}$ denotes the effective channel between the users and the elements working in the connection mode, i.e., $\mathbf{h}_{c}=\mathbf{A}_{a}  \mathbf{h}_{r}$, where $\mathbf{A}_{a} \in \mathbb{R}^{a \times N}$ stands for the channel index matrix of those components connected to the BS. Specifically, $\mathbf{A}_{a}$ is a submatrix of $\mathbf{A}$ consisting of non-zero rows of the mode selection matrix $\mathbf{A}$, satisfying $\mathbf{A}_{a}^{\herm}\mathbf{A}_{a}=\mathbf{A}$.

\subsection{SNR Analysis}
To effectively combine the received signals, the BS employs a low-complexity combining scheme, namely, maximum ratio combining (MRC).  Then, the received signal-to-noise ratio (SNR) is derived as
\begin{align}
    \gamma_{u} =\bar{\gamma} \left({|\mathbf{h}_{d} \!+\!\mathbf{G}^{\herm} (\mathbf{I}\!-\!\mathbf{A}) {\slantedbm{\Phi}} \mathbf{h}_{r}|^2+|\mathbf{A}_{a} \mathbf{h}_{r}|^2}\right), 
\end{align}
where $\bar{\gamma}  \triangleq \frac{p}{{\sigma}^2}$ represents the transmit SNR at the user. To achieve optimal performance, phase shifts and mode selection variables are optimized to adapt varying quality-of-service (QoS) constraints. In specific, the general resource allocation optimization problem can be formulated as
\begin{subequations}
\begin{align}
{\textbf{(P1:)}} \quad  {\mathop {\max }_{{\bm{\theta}},\mathbf{A}}} & \quad  {|\mathbf{h}_{d} \!+\!\mathbf{G}^{\herm} (\mathbf{I}\!-\!\mathbf{A}) {\slantedbm{\Phi}} \mathbf{h}_{r}|^2+|\mathbf{A}_{a} \mathbf{h}_{r}|^2} \\
\qquad \qquad \mathrm{s.t.}~&\quad 
C_{\mathrm{QoS}}^{i} \geq 0, \forall i\in {1,2,...,C_Q},
\end{align}
\end{subequations}
where $C_{\mathrm{QoS}}^{i}$ and $C_Q$ denote the $i$-th QoS constraint and number of QoS constraints, respectively. 

{\color{black}Due to the unit-modulus constraint on the phase shifts, the integer constraints on the mode selection matrix, and the coupling among the optimization variables, problem (P1) is generally non-convex and NP-hard. Similar to the optimization frameworks widely used in RIS-aided systems, the unit-modulus constraints can be handled by techniques such as semidefinite relaxation (SDR), successive convex approximation (SCA), majorization–minimization (MM), or manifold optimization, while the binary mode-selection variables can be addressed via relaxation, penalty-based reformulation, branch-and-bound, or successive rounding strategies. In practice, an alternating optimization framework is often adopted to iteratively update the active beamforming, passive phase shifts, and mode-selection variables, thereby achieving a good balance between complexity and performance. Since the optimization of these variables falls beyond the scope of this paper, interested readers are referred to our previous work for efficient solution algorithms \cite{wang2024radars}. }

Suppose the optimal phase shift vector and mode selection matrix are denoted by ${\bm{\theta}}^{\star}$ and $\mathbf{A}^{\star}$.
Then, the corresponding optimal SNR is denoted as
\begin{align}
    \gamma_{u}^{\mathrm{opt}} =\bar{\gamma} \left({|\mathbf{h}_{d} \!+\!\mathbf{G}^{\herm} (\mathbf{I}\!-\!\mathbf{A}^{\star}) {\slantedbm{\Phi}}^{\star} \mathbf{h}_{r}|^2+|\mathbf{A}_{a}^{\star} \mathbf{h}_{r}|^2}\right), 
\end{align}
where ${\slantedbm{\Phi}}^{\star}$ and $\mathbf{A}_{a}^{\star}$ are derived from ${\bm{\theta}}^{\star}$ and $\mathbf{A}^{\star}$. 
Since $(\mathbf{A}_{a}^{\star})^{\herm}\mathbf{A}_{a}^{\star}=\mathbf{A}^{\star}$, we can rewrite $|\mathbf{A}_{a}^{\star} \mathbf{h}_{r}|^2$ as
\begin{align}
    |\mathbf{A}_{a}^{\star} \mathbf{h}_{r}|^2=\sum\limits_{n=1}^{N}a_n^{\star} \left|[\mathbf{h}_{r}]_i \right|^2,
\end{align}
where $a_n^{\star}$ is the $n$-th diagonal element of $\mathbf{A}$, characterizing the optimal mode selection for $n$-th element in RDARS.
Denote ${a_n^{o}, \forall n \in {1,2,\ldots,N}}$ as the default modes at the RDARS. Then, the optimal SNR can be reformulated as
\begin{align}
    \gamma_{u}^{\mathrm{opt}} 
    & = \bar{\gamma} 
    \Big(
        \underbrace{|\mathbf{h}_{d} 
        + \mathbf{G}^{\herm} (\mathbf{I}-\mathbf{A}^{\star}) 
        {\slantedbm{\Phi}}^{\star} \mathbf{h}_{r}|^2}_{\text{Reflection Gain}}
        + 
        \underbrace{ \sum\nolimits_{n=1}^{N}a_n^{o} \left|[\mathbf{h}_{r}]_i \right|^2}_{\text{distribution gain}} \notag
    \\  & \qquad +
         \underbrace{ \sum\nolimits_{n=1}^{N}(a_n^{\star}-a_n^{o}) \left|[\mathbf{h}_{r}]_i \right|^2}_{\text{Selection Gain}}
    \Big) \label{eq:SNR_decompose}, \\
    &\approx 
    \bar{\gamma} \big(
        G_{\mathrm{RIS}} + G_{\mathrm{DAS}} + G_{\mathrm{Sel}} 
    \big), \label{approx RIS gain}
\end{align}
with $G_{\mathrm{RIS}}\triangleq |\mathbf{h}_{d} 
        + \mathbf{G}^{\herm} {\slantedbm{\Phi}}^{\star} \mathbf{h}_{r}|^2$,  $G_{\mathrm{DAS}}\triangleq\sum\nolimits_{n=1}^{N}a_n^{o} \left|[\mathbf{h}_{r}]_i \right|^2$ and $G_{\mathrm{Sel}}\triangleq \sum\nolimits_{n=1}^{N}(a_n^{\star}-a_n^{o}) \left|[\mathbf{h}_{r}]_i \right|^2$.
The approximation in \eqref{approx RIS gain} holds since $a \ll N$. 
Therefore, the RDARS framework yields superior performance gains compared to conventional RIS and DAS systems, owing to its joint optimization of phase shifts and mode selection.

\subsection{Simulation Results}

In the sequel, we conduct numerical simulations to evaluate the reflection, connection and selection gains of the proposed RDARS architecture. The main basic parameters are configured following the settings in Ref.\inlinecite{wang2024radars}.

\begin{figure}[t]
    \centering
    \subfloat[$N=64$]{
        \includegraphics[width=0.45\textwidth]{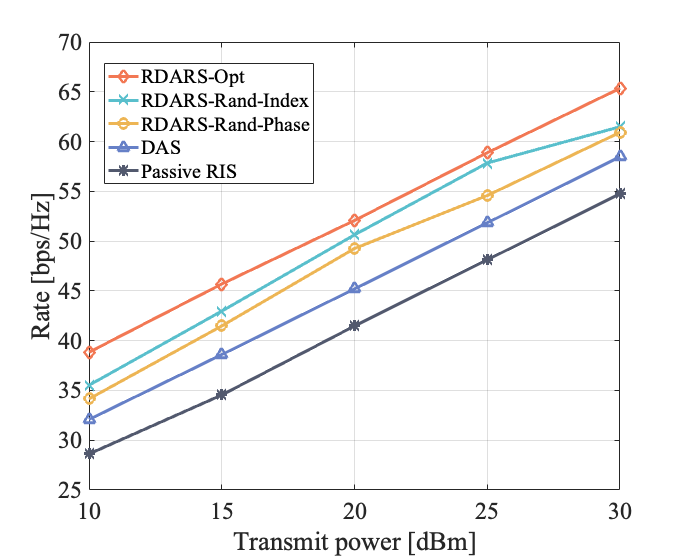}
        \label{fig:n64}
    }
    \hfill
    \subfloat[$N=128$]{
        \includegraphics[width=0.45\textwidth]{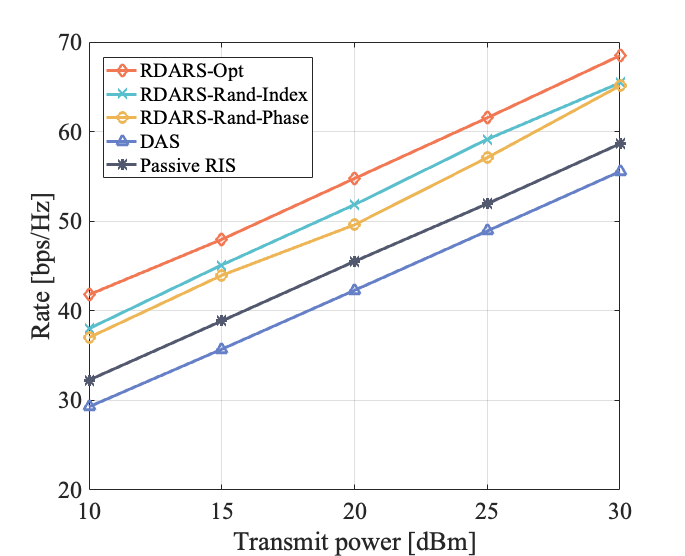}
        \label{fig:n128}
    }
    \caption{\color{black}Rate performance of the RDARS-aided communication system for (a) $N=64$ and (b) $N=128$.}
    \label{Fig.gain} 
\end{figure}

 \begin{figure}[t]
    \centering
    \includegraphics[width=0.45\textwidth]{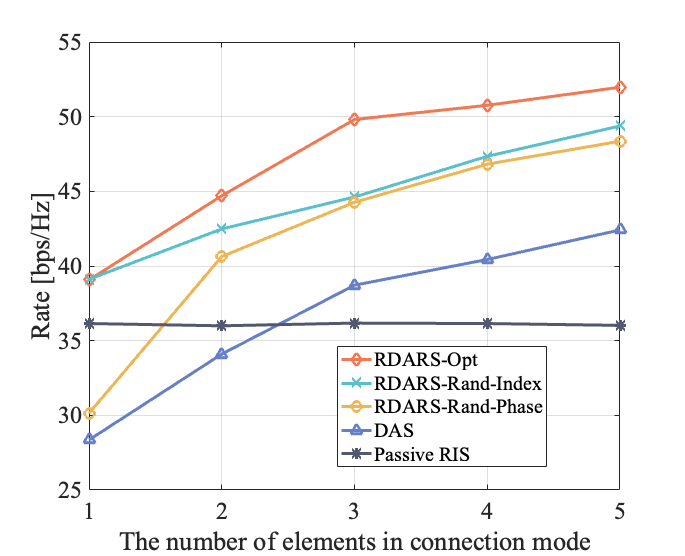}
    \caption{\color{black}Rate gains versus the number of elements in connection mode. }
    \label{Fig.gain_2} 
\end{figure}
{\color{black}
Fig.~\ref{Fig.gain} demonstrates the rate performance of RDARS-aided communication compared to conventional counterparts as the transmit power increases for a varying number of RDARS elements.
Fig.~\ref{Fig.gain} shows that RDARS consistently provides the largest achievable rate across the considered transmit-power range. In particular, RDARS-Opt achieves the best performance for both $N=64$ and $N=128$, which verifies that the joint optimization of mode selection and phase shifts can fully exploit the dynamic mode architecture. Moreover, even when random index selection or random phase design is adopted, RDARS still achieves noticeable gains over DAS and passive RIS, highlighting its robustness under practical low-complexity configurations. It is also observed that enlarging the RDARS size further strengthens these advantages, as the dynamic architecture can more efficiently utilize additional elements than conventional systems.

Fig.~\ref{Fig.gain_2} illustrates the rate performance versus the number of elements in connection mode.
From Fig.~\ref{Fig.gain_2}, it is observed that increasing the number of elements in connection mode generally improves the achievable rate of both RDARS and DAS, whereas the performance of passive RIS remains nearly unchanged. This trend confirms that the elements in connection mode contribute an additional distributed transmission gain. Furthermore, the superiority of RDARS-Opt over RDARS-Rand-Index demonstrates the importance of mode selection, while the gap between RDARS-Opt and RDARS-Rand-Phase reflects the additional gain brought by phase optimization. Therefore, the overall advantage of RDARS is attributed to the combined effect of reflection, distribution, and selection gains, which together enable more efficient and scalable rate enhancement than DAS and passive RIS.}

\section{\MakeUppercase{RDARS-aided Integrated Sensing and Communications}}
ISAC also plays a vital role in 6G networks with the capability of unifying communication and sensing in a shared system. By enabling radar and communication to coexist in the same frequency band and hardware platform, ISAC improves spectrum efficiency, reduces hardware cost, and lowers system complexity. However, higher operating frequencies in ISAC make signals more vulnerable to blockages, leading to significant degradation in both sensing and communication performance. 

In this section, we leverage the RDARS architecture to enhance radar sensing performance by establishing additional reflective and connected links between the BS and the target. The appealing structure of RDARS also mitigates severe signal propagation loss over the four-hop links, as seen in conventional RIS-aided ISAC scenarios.

\subsection{System Model}
\begin{figure}[t]
    \centering
    \includegraphics[width=0.45\textwidth]{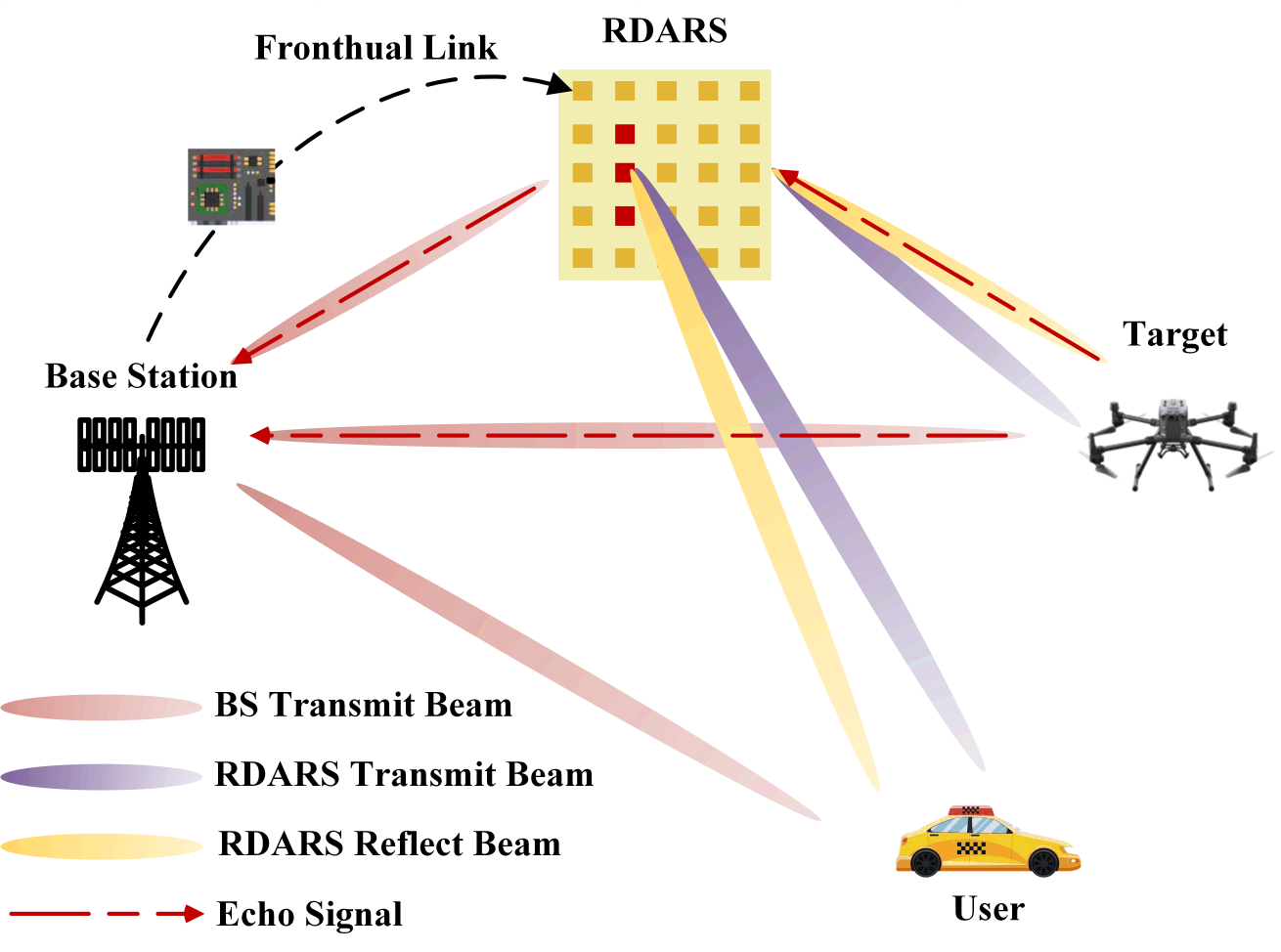}
    \caption{Illustration of RDARS-aided ISAC transmission.}
    \label{Fig.ISAC_Scenario} 
\end{figure}

As shown in Fig.~\ref{Fig.ISAC_Scenario}, we consider a RDARS-aided ISAC system with one BS and one RDARS serving a single-antenna user and a target.\footnote{\color{black}We consider a simplified baseline ISAC model with a single user and a quasi-static far-field channel, so as to highlight the essential sensing gain mechanism enabled by RDARS. More general scenarios, such as multi-user systems, multi-antenna transceivers, time-varying channels, and near-field propagation, are beyond the scope of this paper. In particular, multi-user and multi-antenna RDARS-aided designs are discussed in \inlinecite{zhang2025radars}.} The BS has $M$ transmit and $M$ receive antennas, and the RDARS follows the configuration in Section.~\ref{sec:rdars_aided_communications}. Note that the elements of RDARS in connection mode perform downlink transmission to enhance sensing. 
Under the quasi-static far-field channel environment,  the channels from  the BS and RDARS to the communication user are denoted as $\mathbf{u}_{d} \in\mathbb{C}^{M \times 1}$ and $\mathbf{u}_{r} \in \mathbb{C}^{N \times 1}$, respectively. Similarly, the channels from the BS and RDARS to the target are denoted as $\mathbf{t}_{d}\in\mathbb{C}^{M \times 1}$ and $\mathbf{t}_{r}\in \mathbb{C}^{N \times 1}$, respectively.  Besides, the channel between the BS and the RDARS is still denoted as $\mathbf{G} \in \mathbb{C}^{N \times M}$.

Denote the transmit ISAC data symbol as $s$ with $\mathbb{E}[{{s}} {{s}}^*]=1$ and the transmit beamforming vectors for the BS antennas and elements in connection mode as $\mathbf{f}_{b}\in\mathbb{C}^{M \times 1}$ and $\mathbf{f}_{r}\in\mathbb{C}^{a \times 1}$, respectively. Then, the received signal at the communication user is denoted as
\begin{equation}
  {y}_{c}
  = 
  {\underbrace{\begin{bmatrix} \mathbf{u}_{d} \!+\!\mathbf{G}^{\herm} (\mathbf{I}\!-\!\mathbf{A}) {\slantedbm{\Phi}} \mathbf{u}_{r} \\   \mathbf{A}_{a} \mathbf{u}_{r}   \end{bmatrix}   }_{ \mathbf{u} }}^{\herm} \underbrace{\begin{bmatrix} \mathbf{f}_{b} \\ \mathbf{f}_{r}\end{bmatrix}}_{\mathbf{f}} {{s}} + n_{c},
\end{equation} 
with $\mathbf{u}$ representing the effective channels. $n_{c}$ represent the AWGN obeying the CSCG distributions with zero mean and variance of ${\sigma_{b}^2}$ and $\mathbf{f}$ stands for the total beamforming vector with power budget of $|\mathbf{f}|^2=p$. Therefore, the received SNR is derived as $\gamma_{c} = \frac{|\mathbf{u}^{\herm}\mathbf{f}|^2}{\sigma^2}$.

{\color{black}
\subsection{Radar SNR Analysis}
For the sensing target, the received echo signal at the BS is denoted as
\begin{align}
    \mathbf{y}_{s}
  \!=\! \alpha ( \mathbf{t}_{d} \!+\!\mathbf{G}^{\herm} (\mathbf{I}\!\!-\!\!\mathbf{A}) {\slantedbm{\Phi}} \mathbf{t}_{r} ) { \underbrace{\begin{bmatrix} \mathbf{t}_{d} \!+\!\mathbf{G}^{\herm} (\mathbf{I}\!\!-\!\!\mathbf{A}) {\slantedbm{\Phi}} \mathbf{t}_{r} \\   \mathbf{A}_{a} \mathbf{t}_{r}  \end{bmatrix}   }_{ \mathbf{t} } }^{\herm} \begin{bmatrix} \mathbf{f}_{b} \\ \mathbf{f}_{r}\end{bmatrix} {{s}} \!+\!  \mathbf{n}_{s},
\end{align}
where $\alpha$ refers to the radar cross-section coefficient and $\mathbf{n}_{s}$ is the AWGN vector subject to $\mathbf{n}_{s} \sim \mathcal{CN}(0, \sigma^2 \mathbf{I}_{M})$.

With the receiver filter $\mathbf{w}\in\mathbb{C}^{M \times 1}$, the radar SNR is derived as
\begin{align}
\gamma_{s} =  \frac{\alpha^2|\mathbf{w}^{\herm}\mathbf{t}_{b}|^2 |\mathbf{t}_{b}^{\herm}\mathbf{f}_{b}+\mathbf{t}_{c}^{\herm}\mathbf{f}_{r}|^2}{\sigma^2|\mathbf{w}|^2},
\end{align}
where $\mathbf{t}_{b}=\mathbf{t}_{d} \!+\!\mathbf{G}^{\herm} (\mathbf{I}\!-\!\mathbf{A}) {\slantedbm{\Phi}} \mathbf{t}_{r} $ and $\mathbf{t}_{c}=\mathbf{A}_{a} \mathbf{t}_{r} $ representing the effective channels between the BS and the target related to reflective and connection links, respectively.
Generally, the radar SNR maximization problem can be formulated as 
\begin{subequations}
\begin{align}
{\textbf{(P2:)}} \quad  {\mathop {\max }_{ {\bm{\theta}},\mathbf{A} , \mathbf{w}, \mathbf{f} }} & \qquad  \gamma_{s} \\
\qquad \qquad \mathrm{s.t.}~&\quad 
S_{\mathrm{ISAC}}^{i} \geq 0, \forall i\in {1,2,...,S_Q},
\end{align}
\end{subequations}
where $S_{\mathrm{ISAC}}^{i}$ and $S_Q$ denote the $i$-th ISAC constraint and number of ISAC constraints, respectively.

For simplicity, we assume the BS employs MRC for the received echo signal and maximum ratio transmission (MRT) for signal transmission. Thus, the radar SNR can be further simplified as
\begin{align}
\gamma_{s} = & 
{\bar{\gamma}}{\alpha^2}  \left(|\mathbf{t}_{d} \!+\!\mathbf{G}^{\herm} (\mathbf{I}\!-\!\mathbf{A}) {\slantedbm{\Phi}} \mathbf{t}_{r}|^2+|\mathbf{A}_{a} \mathbf{t}_{r}|^2\right) \notag \\
&\qquad |\mathbf{t}_{d} \!+\!\mathbf{G}^{\herm} (\mathbf{I}\!-\!\mathbf{A}) {\slantedbm{\Phi}} \mathbf{t}_{r}|^2, \\
=&{\bar{\gamma}}{\alpha^2} (|\mathbf{t}_{d} \!+\!\mathbf{G}^{\herm} (\mathbf{I}\!-\!\mathbf{A}) {\slantedbm{\Phi}} \mathbf{t}_{r}|^4  \notag\\
& + \sum\limits_{n=1}^{N}a_n \left|[\mathbf{t}_{r}]_i \right|^2 |\mathbf{t}_{d} \!+\!\mathbf{G}^{\herm} (\mathbf{I}\!-\!\mathbf{A}) {\slantedbm{\Phi}} \mathbf{t}_{r}|^2 ).
\end{align}
In contrast to conventional RIS-aided ISAC systems, the proposed RDARS architecture yields significantly higher radar SNR. This is attributed to the RDARS connection mode, which establishs additional two-hop signal paths, thereby providing a diversity gain over purely passive RIS counterparts.
}

\subsection{Simulation Results}

\begin{figure}[t]
    \centering
    \subfloat[$a=2$]{
        \includegraphics[width=0.45\textwidth]{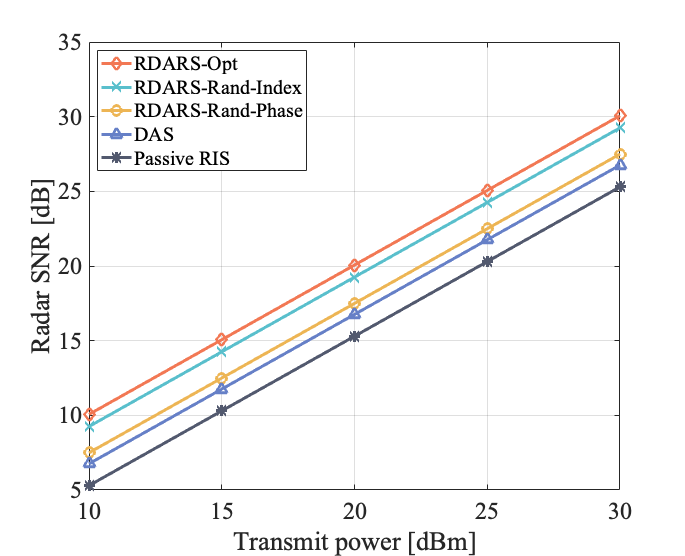}
        \label{fig:a2}
    }\hfill 
    \subfloat[$a=4$]{
        \includegraphics[width=0.45\textwidth]{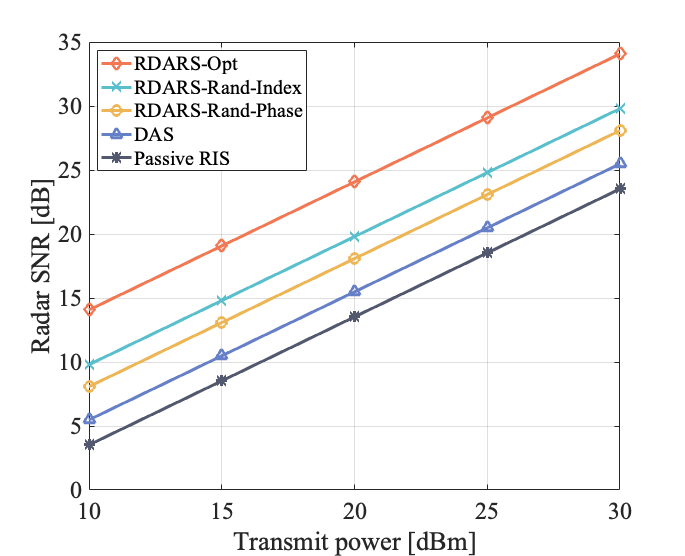}
        \label{fig:a4}
    }
    \caption{\color{black}Radar SNR gains for RDARS-aided ISAC system when (a) $a=2$ and (b) $a=4$.}
    \label{Fig.radargain}
\end{figure}

Simulations are conducted to verify the advantage of RDARS architecture in ISAC scenarios, as shown in Fig.~\ref{Fig.radargain}. The phase shift coefficients and mode selection matrix are optimized by the penalty-based iterative algorithm proposed in \cite{zhang2025radars} to maximize the radar SNR while satisfying the QoS constraint of the communication user, i.e., $\gamma_{c} \geq \gamma_{\mathrm{th}}$ with $\gamma_{\mathrm{th}}$ denoting the minimum communication SNR requirement. 
We set $N=64$ total elements in RDARS. The remaining basic parameters are configured following the settings in Ref.\inlinecite{zhang2025radars}.



{\color{black}
Fig.~\ref{Fig.radargain} shows the radar SNR versus the transmit power for a varying number of elements in connection mode. It can be observed that the radar SNR of all schemes increases with the transmit power for both $a=2$ and $a=4$. RDARS-Opt consistently achieves the best performance across the entire power range, while RDARS-Rand-Index and RDARS-Rand-Phase also outperform DAS and passive RIS, demonstrating the robustness of RDARS even under suboptimal configurations.
Moreover, increasing the number of connection-mode elements from $a=2$ to $a=4$ further improves the radar SNR, which confirms the additional sensing gain introduced by the connection functionality. In particular, the gap between RDARS-Opt and RDARS-Rand-Index reflects the gain brought by mode selection, whereas the gap between RDARS-Rand-Index and RDARS-Rand-Phase captures the additional benefit of phase optimization. These results verify that the sensing advantage of RDARS originates from the joint effect of reflection gain, distribution gain, and selection gain, enabling more efficient radar SNR enhancement than DAS and passive RIS.}

\section{\MakeUppercase{Challenges and Future Directions}}
Despite the promising potential of RDARS architecture, several challenges remain before its large-scale deployment in future wireless systems. In this section, we will elaborate on the major open issues and outlines possible research directions from three key aspects: hardware and implementation limitations, resource allocation and optimization, and cell-free RDARS architectures.

\subsection{Practical Implementation Issues}
Although the theoretical benefits of RDARS have been extensively demonstrated, translating these advantages into real-world systems remains non-trivial due to hardware imperfections and system-level constraints. 

\subsubsection{Hardware Non-idealities}
In practice, the components within RDARSs, such as phase shifters, amplifiers, and signal converters, are subject to finite resolution, nonlinearity, and thermal noise \cite{Wang2023HWI}. The finite-resolution phase quantization inevitably introduces phase errors, leading to beam misalignment and degraded radar SNR and communication rate. Furthermore, amplifier noise and gain fluctuations can distort both the transmitted and reflected signals, reducing the sensing precision and reliability of the communication link. 

\subsubsection{Circuit Manufacturing and Coupling Effects}
Due to the complex circuit structure of RDARS elements that integrate both reflection and connection functionalities, non-ideal manufacturing processes can cause energy leakage and inter-element coupling. The coupling between the connected and reflected units alters the intended phase response, leading to undesired signal interference. These coupling effects are particularly pronounced when the elements are densely packed or when the hardware operates at high frequencies (e.g., millimeter-wave or THz bands). Accurate modeling and compensation for these effects are therefore essential for high-fidelity system design.

To address these challenges, future RDARS designs should incorporate hardware–algorithm co-design,  and circuit-level calibration. Integrating advanced electromagnetic co-simulation with signal optimization algorithms can enable more accurate and efficient performance prediction, facilitating the transition from theoretical analysis to hardware prototyping.

\subsection{Resource Allocation and Intelligent Optimization}
The optimization of RDARS systems is inherently complex due to the joint design of multiple parameters, including the phase shifts, mode selection, and power allocation. This results in a mixed-integer and highly nonconvex optimization problem, whose global optimum is extremely difficult to obtain in closed form. Traditional optimization methods based on convex approximation or alternating optimization often lead to high computational complexity and suboptimal performance, particularly in large-scale scenarios. 

\subsubsection{Challenge of Mixed-Integer Programming}
The discrete nature of mode selection combined with continuous phase shift variables makes the resource allocation problem combinatorial. The computational cost grows exponentially with the number of elements, making exact optimization impractical for real-time or large-scale systems. Developing efficient and scalable algorithms that can balance optimality and computational feasibility remains a central challenge. 

\subsubsection{Potential of Deep Learning-Based Optimization}
The powerful nonlinear approximation and pattern recognition capabilities of deep learning (DL) offer promising opportunities for addressing these complex optimization problems. By learning the mapping between channel states and optimal resource configurations, DL-based approaches can provide near-optimal solutions with much lower computation time once trained. Furthermore, deep reinforcement learning (DRL) and graph neural networks (GNNs) are emerging as effective tools for distributed optimization and mode selection in complex network topologies \cite{10697115}. 



\subsection{Toward Cell-Free RDARS Networks}
Inspired by the concept of cell-free massive MIMO, where distributed access points (APs) cooperatively serve users without predefined cell boundaries, cell-free RDARS networks aim to extend such cooperation to sensing and communication simultaneously. 

\subsubsection{Cooperative Gain and Interference Management}
In cell-free RDARS architectures, multiple distributed RDARSs can collaboratively enhance the radar sensing accuracy and communication reliability through coherent beamforming. However, coordinating the transmission and reflection behaviors across geographically distributed RDARSs introduces new challenges in interference control, synchronization, and joint signal processing. Designing cooperative beamforming and interference alignment strategies that account for both sensing and communication metrics is a critical research direction. 

\subsubsection{Synchronization and Channel Estimation}
Accurate synchronization among RDARSs is essential for coherent operation, especially when combining reflected and connected signals from different devices. Time and phase synchronization errors can accumulate, causing destructive interference in joint radar–communication operations. In addition, efficient channel estimation across multiple RDARSs and users becomes increasingly complex due to the high-dimensional composite channels and dynamic environments. Future research could explore distributed synchronization protocols, pilot design, and low-overhead channel estimation methods to maintain coordination among multiple RDARSs. 


By enabling global cooperation and eliminating cell boundaries, cell-free RDARS networks are expected to provide seamless coverage, enhanced sensing accuracy, and improved energy efficiency. However, realizing these benefits requires overcoming significant design and coordination challenges at both the hardware and algorithmic levels.

\section{\MakeUppercase{Conclusion}}
In this article, we introduced RDARS as a transformative evolution of conventional RIS technology. By integrating dynamically reconfigurable connection and reflection elements, RDARS simultaneously harnesses reflection, connection, and selection gains, enabling highly flexible, energy-efficient, and reliable wireless coverage. 
We highlighted its advantages in terms of its operating principle, potential gains, deployment aspects, and comparisons with related structures. 
SNR analysis was performed for both RDARS-aided uplink communication and ISAC systems to theoretically and experimentally demonstrate the performance gains of RDARS over existing benchmark schemes.

While the RDARS structure hold substantial advantages for future 6G networks, several open challenges had been discussed. Addressing hardware imperfections and fronthaul limitations, developing intelligent and scalable resource allocation algorithms, and extending RDARS technology to cell-free cooperative networks represent the three most critical research frontiers. Future efforts should emphasize cross-layer design, hardware–algorithm co-optimization, and AI-driven adaptation to fully unleash the potential of RDARSs. By integrating these innovations, RDARS architecture is poised to become a new paradigm towards reconfigurable environment for future communications.


\bibliography{JCIN-RDARS-Final}

\end{document}